\documentclass[pra,twocolumn]{revtex4-2}
\usepackage{amssymb}
\usepackage{amsfonts}
\usepackage{amsmath}
\usepackage{graphicx}
\usepackage{bm}
\usepackage{braket}
\usepackage{epstopdf}
\usepackage{xcolor}

\begin{document}

\title{Tightly self-trapped modes and vortices in three-dimensional bosonic
condensates with the electromagnetically-induced gravity}
\author{Zibin Zhao$^{1}$}
\thanks{These authors contributed equally to this work.}
\author{Guilong Li$^{1}$}
\thanks{These authors contributed equally to this work.}
\author{Huanbo Luo$^{1,2}$}
\email{huanboluo@fosu.edu.cn}
\author{Bin Liu$^{1,3}$}
\author{Gui-hua Chen$^4$}
\email{cghphys@gmail.com}
\author{Boris A. Malomed$^{5,6}$}
\author{Yongyao Li$^{1,3}$ }
\affiliation{$^1$School of Physics and Optoelectronic Engineering, Foshan University,
Foshan 528000, China\\
$^2$Department of Physics, South China University of Technology, Guangzhou
510640, China\\
$^3$Guangdong-Hong Kong-Macao Joint Laboratory for Intelligent Micro-Nano
Optoelectronic Technology, Foshan University, Foshan 528225, China\\
$^{4}$School of Electronic Engineering \& Intelligentization, Dongguan
University of Technology, Dongguan 523808, China\\
$^5$Department of Physical Electronics, School of Electrical Engineering,
Faculty of Engineering, Tel Aviv University, Tel Aviv 69978, Israel\\
$^6$Instituto de Alta Investigaci\'{o}n, Universidad de Tarapac\'{a},
Casilla 7D, Arica, Chile }

\begin{abstract}
The $1/r$ long-range interaction, induced by laser illumination, offers a
mechanism for the implementation of stable self-trapping in Bose-Einstein
condensates (BECs) in the three-dimensional free space. Using the
variational approximation and numerical solutions, we find that self-trapped
states in this setting
, with attractive nonlocal and repulsive local interactions, resemble tightly-bound compactons.
However, these are not true compactons but rather
\textit{tightly self-trapped modes} (TSTMs), with small-amplitude nonvanishing tails.
The structure of the self-trapped states is explained by an analytical
solution for their tails. Further, we demonstrate that stable %
TSTMs with embedded vorticity, exist in the same setting,
with winding numbers up to $S=6$ (at least). Addressing two-TSTM
interactions, we find that pairs of ground states (GSs, with $S=0$), as well
as vortex-vortex and vortex-antivortex pairs (with $S_1=S_2$ and $S_1=-S_2$,
respectively), form stably rotating bound states. Head-on collisions between
vortex TSTMs, set in slow motion by kicks, are inelastic, resulting in their
merger into a GS soliton, that may either remain at the collision position
or move aside, shedding the angular momentum with emitted
radiation, or, alternatively, lead to the formation of a vortex that also
moves aside.
\end{abstract}

\maketitle


\section{Introduction}

In the realm of nonlinear physics, solitons are known for their unique
stability and propagation characteristics \cite%
{zabuskyInteraction1965,Shabat,Zakharov,ablowitzSolitons1981,Newell,drazinSoliton1989}%
. Solitons exist in a great variety of models \cite%
{scottThesoliton1973,friedbergClass1976,hirotaSoliton1981,hirotaSoliton1989,
kivsharDynamics1989,Abdullaev,drummond1998,Mollenauer,serkinNovel2000,Hulet,
Khayk,helalSoliton2002,KA,Peyrard,kevrekidis2016,CQDLVector-soliton,CQmutiwavesoliton, LZengmoirelattice,malomed2022}%
, where they typically exhibit exponentially decaying tails. As a class of
tailless self-trapped states with a compact support, Rosenau and Hyman
(1993) introduced the concept of compactons \cite{rosenau1993compactons},
which are produced by real nonlinear equations. A typical structure of the compacton
solution is
\begin{equation}
u(r)=\left\{
\begin{array}{c}
A\cos ^{2}\left( \pi r/D\right) H_{\mathrm{s}}\left( D/2-r\right) ,~\mathrm{%
at}~r<D/2, \\
0,~\mathrm{at}~r>D/2,%
\end{array}%
~\right.  \label{compacton}
\end{equation}%
where $r$ is the radial coordinate, $H_{\mathrm{s}}$ is the Heaviside
step-function, while $A$ and $D$ denote the amplitude and diameter of the
spatially confined mode \cite{rosenau2018compactons}.

Compactons are also known as solutions of specially designed discrete models
\cite{Kivshar,Kivshar2}. In particular, stable single- and multisite
discrete compactons can be maintained by the combination of a deep optical
lattice and fast time-periodic modulation of the nonlinearity \cite%
{abdullaev2010compactons}. However, compacton-like modes have not been
previously predicted in continuous three-dimensional (3D) BEC systems
(without the use of deep optical lattices). The objective of the present
work is to establish stable tightly self-trapped modes
(TSTMs) as natural ground and excited (vortical) states in a BEC with
specific long-range inter-atomic interactions.

It is well known that long-range interactions offer a wide range of
possibilities for the creation of spatial modes in BEC. Common types of such
interactions are represented by dipole-dipole forces, which have been
studied in \ detail theoretically and experimentally \cite%
{santosBose2000,kgralBose2000,URfischer2006,baranovTheoretical2008,Tikho,Lewenstein, chomazDipolar2022,defenuLong2023,bigagliObservation2024}%
, and soft-core phenomena. Anisotropic dipole-dipole interactions, with the
interatomic potential $\sim 1/r^{3}$, occur in BECs of atoms with large
magnetic moments. Soft-core interactions, on the other hand, arise from the
coupling between Rydberg and ground atomic states, with widely tunable
strengths \cite{mohapatraCoherent2007,johnson2008,heidemann2008rdbderg,
saffmanQuantum2010,balewskiRydberg2024}. Unlike hard-core interactions,
where the potential becomes infinite at short distances, the soft-core
potential saturate to a finite value as the atoms approach each other. These
two types of long-range interactions not only support the formation of
self-trapped 3D states, but also enable the emergence of novel quantum
states of matter, such as supersolids \cite%
{zhanglong2021,liTwo2024,zhangPhase2021,heQuantum2024,yangTwo2024,zhangSupersolidity2019, chenGiant2023,li2024,zhuThree2024}%
. 

In addition to these settings, it was proposed by O'Dell et al. \cite%
{o2000bose} to induce an atomic interaction with an attractive potential
similar to the gravitational one, scaling as $-1/r$, by means of a set of
six suitably arranged far-off-resonant laser beams. An estimate demonstrates
that this interaction may be stronger than the proper gravitational
attraction between the same atoms by $\simeq 17$ orders of magnitude,
offering an efficient method to simulate quasi-gravitational effects in
quantum systems, such as the realization of self-trapping in the free space
\cite{papadopoulosBose2007,lushnikov2002,maucherRotating2010}, or
investigate quantum nonlinear phenomena in the number-conserving analogues
of gravity within BEC \cite{URfischer2024}.

For the analysis of ground and excited states in BEC\ with long-range
interactions, an extended variational approximation (VA) was elaborated,
based on an ansatz in the form of a sum of Gaussians with a common center
but different width parameters \cite%
{rauVariationalI2010,rauVariationalII2010}, the commonly known VA based on
the single Gaussian being the lowest-order form of this approach. The
VA-predicted results become essentially more accurate (while the
computational procedure becomes more cumbersome) with the increase of the
number of the Gaussian terms in the ansatz.

In this paper, we develop VA for TSTMs, using a compacton ansatz, and
compare the ensuing results with numerical findings, as well as with VA
results produced by a simple Gaussian ansatz. We conclude that the results
for the TSTMs, generated by the compacton ansatz, are essentially more
accurate (closer to the numerical findings) than their Gaussian-based
counterparts.
In addition to the ground state (GS) of TSTMs, the analysis
reveals stable TSTMs with embedded vorticity. It is well known that, in
other models admitting vortex solitons, those carrying high topological
charges are usually unstable in the free space against azimuthal
perturbations, which break the vortices into fragments or lead to their
decay into GS \cite{malomed2022,kartashovFrontiers2019}. In the present
situation, we find that the gravity-like self-attraction stabilize vortex
states with topological charges up to (at least) $S=6$. The fact that the
long-range attraction with the potential $\sim -1/r$ can support stable
vortex modes (in particular, ones with high values of $S$) has not been
reported previously. In this connection, it is relevant to stress that,
while many mechanisms were proposed for the stabilization of vortex
solitons, the stability remains a challenging issue for $S>1$, especially in
the 3D space \cite%
{quirogaInternal1999,towers2001,quirogaStable1997,desytnikovThree2000,Barcelona, Sakaguchi,HanPu,dongVortex2022,liuHigher2023,dongStable2023,zhangHigher2019, zhangStabilization2022,dongStableHigher2023,listable2024,wangRydberg2024,xuVortex2024}%
.

The following material is organized as follows. The 3D model with the
gravity-like self-attraction is introduced in Section 2. In that section, we
also produce an analytical solution for the TSTMs'
asymptotic tails. It demonstrates that the edge of the self-trapped state
becomes much sharper for the states with a sufficiently large norm, which
explains the proximity of their shape to the compacton. VA for the
TSTMs of the GS type (with zero vorticity) is elaborated and compared 
to numerical findings in Section 3. Numerical results for 3D stationary 
TSTMs, of both the GS
and vortex types, including the test of their stability, are reported in
Section 4. Further, dynamics of the GSs and vortices, including the
formation of stable orbiting bound states and collisions between moving %
TSTMs, is investigated by means of direct simulations in
Section 5. In particular, it is demonstrated that GS pairs, as well as
vortex-antivortex ones, form stable configurations with mutual orbiting,
while collisions result in inelastic outcomes. The paper is concluded by
Section 6.

\section{The model}

The potential of the gravity-like inter-atomic attraction for BEC\
illuminated by the appropriate set of laser beams with wave vector $k$ and
intensity $I$ was derived in Ref. \cite{o2000bose}:
\begin{equation}
U\left( R\right) =-\frac{11}{4\pi }\frac{Ik^{2}\alpha ^{2}}{c\epsilon
_{0}^{2}}\frac{1}{R}=-\frac{K}{R},  \label{1-rpotential}
\end{equation}%
where $c$ and $\epsilon _{0}$ are the light speed and permittivity in the
free space, $\alpha (k)$ is the isotropic dynamical polarizability of the
laser, and $I$ is the intensity of the laser, and $R$ is the distance
between the interacting atoms. The setup is designed so that the usual
dipole-dipole-like term, $\sim R^{-3}$ is averaged out, while the one $\sim
-R^{-1}$ does not vanish. Thus, BEC with the gravity-like attraction is
governed by the Gross-Pitaevskii equation (GPE), written as
\begin{equation}
i\hbar \frac{\partial }{\partial T}\Psi =-\frac{\hbar ^{2}}{2M}\nabla
^{2}\Psi +G|\Psi |^{2}\Psi -K\Psi \int \frac{|\Psi (\mathbf{R^{\prime }}%
)|^{2}}{|\mathbf{R}-\mathbf{R^{\prime }}|}d\mathbf{R^{\prime }},  \label{GPE}
\end{equation}%
where $G=4\pi \hbar ^{2}a_{s}/M$ is the coefficient of the contact
interaction ($a_{s}$ is the \textit{s}-wave scattering length of atomic
collisions, and $M$ is the atomic mass).\ The estimate for $K$ given in Ref.
\cite{o2000bose} for a $\mathrm{CO_{2}}$ laser light with intensity $I=100\
\mathrm{MW}/\mathrm{cm}^{2}$ amounts to $-K/R\approx -2\times 10^{-15}\
\mathrm{eV}$ at $R=100\ \mathrm{nm}$. The total number of atoms in this
setup is
\begin{equation}
\mathcal{N}=\int |\Psi (\mathbf{R})|^{2}d^{3}\mathbf{R},
\label{total-atom-number}
\end{equation}%
the set of control parameters being $a_{s},I$ and $\mathcal{N}$. Using the
normalization
\begin{equation}
T=t_{0}\cdot t,\quad \mathbf{R}=l_{0}\cdot \mathbf{r},\quad \Psi =%
l_{0}^{-\frac{3}{2}}\cdot \psi ,
\end{equation}%
where $t_{0}$ and $l_{0}=\sqrt{\hbar t_{0}/M}$ are time and length scales,
the dimensionless GPE can be obtained as:
\begin{equation}
i\partial _{t}\psi =-\frac{1}{2}\nabla ^{2}\psi +g|\psi |^{2}\psi -\kappa
\psi \int \frac{|\psi (\mathbf{r^{\prime }})|^{2}}{|\mathbf{r}-\mathbf{%
r^{\prime }}|}d\mathbf{r^{\prime }},  \label{dime-less-GPE}
\end{equation}%
where
\begin{equation}
g=4\pi a_{s}/l_{0},\quad \kappa =KMl_{0}/\hbar ^{2}
\end{equation}%
are dimensionless strengths of the local and nonlocal interactions,
respectively.

Equation (\ref{GPE}) shares the property of the Galilean invariance with the
broad class of nonlinear Schr\"{o}dinger equations, which include the same
linear operator. This means that the substitution of%
\begin{equation}
\psi \left( \mathbf{r},t\right) =\exp \left( i\mathbf{v}\cdot \mathbf{r}%
+i\left( v^{2}/2\right) t\right) \tilde{\psi}\left( \mathbf{r}-\mathbf{v}%
t,t\right) ,  \label{tilde}
\end{equation}%
with an arbitrary vectorial velocity $\mathbf{v}$, in Eq. (\ref%
{dime-less-GPE}) transforms it into the same equation, but with $\psi $
replaced by $\tilde{\psi}$ and $\mathbf{r}$ replaced by $\mathbf{\tilde{r}%
\equiv r}-\mathbf{v}t$. In other words, the Galilean transform (\ref{tilde})
converts any solution of Eq. (\ref{dime-less-GPE}) into its counterpart
moving with velocity $\mathbf{v}$ and carrying additional phase $\mathbf{v}%
\cdot \mathbf{r}+\left( v^{2}/2\right) t$.

In fact, Eq. (\ref{dime-less-GPE}) admits additional rescaling (not used
here), which makes it possible to set $g=1$ (unless $g=0$)
and $\kappa =1$ \cite{NavarreteSpatial}. In the subsequent numerical studies,
we refer to $^{87}\mathrm{Rb}$ and select $l_{0}=0.5$ $\mathrm{\mu }$m
(which corresponds to $t_{0}\approx 0.03$ ms), without using this rescaling.

It is relevant to mention that Eq. (\ref{dime-less-GPE}) belongs to the
class of Schr\"{o}dinger-Poisson (SP) equations, which were considered in
the context of various physical models -- in particular, cosmology %
\cite{Penrose,Ruiz,Engels,Witten,HYSchiveCosmic,LhuiWaveDark}. The
existence of GS solutions to SP equations in 3D was rigorously established
(in particular, by means of the variational method
\cite{Schr-Poisson variational,Milan,RuffintGS-SP,
RHChavanisGS-SP,PHchavanisMRrelation}). Dynamics of delocalized vortices in
plasmas, governed the SP system, was addressed too \cite{Shukla}. Vortex
solitons produced by the SP and their stability against the spontaneous
splitting in 2D and 3D were also studied in detail 
\cite{Asakawa2024,Nikolaieva2021,Michinel,Lashkin3D,Nikolaieva,ASDmitriev,YONikolaieva}%
, as well as incoherent (turbulent) self-trapped states \cite{Garnier}.

Stationary solutions to Eq. (\ref{dime-less-GPE}) with chemical potential $%
\mu $ are looked for as
\begin{equation}
\psi =\varphi (\mathbf{r})\exp (-i\mu t),  \label{stationary-solution}
\end{equation}%
with (generally speaking, complex) function $\varphi $ obeying the
stationary GPE:
\begin{equation}
\mu \varphi =-\frac{1}{2}\nabla ^{2}\varphi +g|\varphi |^{2}\varphi -\kappa
\varphi \int \frac{|\varphi (\mathbf{r^{\prime }})|^{2}}{|\mathbf{r}-\mathbf{%
r^{\prime }}|}d\mathbf{r^{\prime }}.  \label{stationary-GP}
\end{equation}%
The energy (Hamiltonian) of GPE (\ref{dime-less-GPE}) is
\begin{equation}
\begin{split}
E=& \int d\mathbf{r}\left[ \frac{1}{2}|\nabla \varphi |^{2}+\frac{1}{2}%
g|\varphi |^{4}\right] \\
& -\frac{\kappa }{2}\iint \frac{d\mathbf{r}d\mathbf{r^{\prime }}}{|\mathbf{r}%
-\mathbf{r^{\prime }}|}|\varphi (\mathbf{r^{\prime }})|^{2}|\varphi (\mathbf{%
r})|^{2}.
\end{split}
\label{energy}
\end{equation}

It is relevant to analyze the tail of self-trapped states produced by Eq. (%
\ref{stationary-GP}) with $\mu <0$ at $r\rightarrow \infty $. In this limit,
the linearization of Eq. (\ref{stationary-GP}) takes the following form, in
the lowest approximation:%
\begin{equation}
\mu \varphi +\frac{1}{2}\left( \frac{d^{2}}{dr^{2}}+\frac{2}{r}\frac{d}{dr}%
\right) \varphi +\kappa \frac{N}{r}\varphi =0,  \label{linear}
\end{equation}%
where
\begin{equation}
N=\int |\varphi|^2 d^3\mathbf{r}  \label{total-norm}
\end{equation}
is the total norm of the self-trapped state (the same asymptotic equation (%
\ref{linear}) is relevant for the self-trapped states with embedded
vorticity). A straightforward asymptotic solution of Eq. (\ref{linear}) is%
\begin{equation}
\varphi =\varphi _{0}r^{\kappa N/\sqrt{-2\mu }-1}\exp \left( -\sqrt{-2\mu }%
r\right) ~\mathrm{at}~r\rightarrow \infty ,  \label{asympt}
\end{equation}%
where $\varphi _{0}$ is a constant (this solution is different from the
commonly known ground-state wave function of the hydrogen atom, which
formally corresponds to $\kappa N/\sqrt{-2\mu }-1=0$). In the case of $N\gg
\sqrt{-2\mu}/\kappa,$ the gravity-like attraction draws the tail into the
body of the self-trapped state, thus making the edge of the state sharper,
which is the distinct feature of TSTMs. Indeed, the
waveform (\ref{asympt}) with large $N$ has a maximum at $r_{\max }\approx
\kappa N/\left( -2\mu \right) $, which, roughly speaking, determines the
QC's radius, while the maximum does not exists at\ $N<\sqrt{-2\mu }/\kappa $.

\section{The variational approximation for the GS (ground state)}

The GS wave function $\varphi $ is real, hence Eq. (\ref{stationary-GP})
amounts to the real one:
\begin{equation}
\mu \phi =-\frac{1}{2}\left( \frac{d^{2}}{dr^{2}}+\frac{2}{r}\frac{d}{dr}%
\right) \phi -\kappa \phi (r)\int \frac{\phi ^{2}(\mathbf{r}^{\prime })}{%
\left\vert \mathbf{r}-\mathbf{r}^{\prime }\right\vert }d\mathbf{r}^{\prime
}+g\phi ^{3}.  \label{GP}
\end{equation}%
The Lagrangian corresponding to Eq. (\ref{GP}) is
\begin{equation}
\begin{split}
L=& \int d\mathbf{r}\left[ \frac{1}{2}\left( \frac{d\phi }{dr}\right) ^{2}-%
\frac{\mu }{2}\phi ^{2}+\frac{g}{4}\phi ^{4}\right] \\
& -\frac{\kappa }{2}\iint \frac{d\mathbf{r}d\mathbf{r^{\prime }}}{|\mathbf{r}%
-\mathbf{r^{\prime }}|}\phi ^{2}(\mathbf{r})\phi ^{2}(\mathbf{r^{\prime }}).
\end{split}
\label{Lagrange}
\end{equation}%
Two VA \textit{ans\"{a}tze} for GS in 3D are adopted as the compacton $\phi
_{1}(r)$ and Gaussian $\phi _{2}(r)$, \textit{viz}.,
\begin{equation}
\phi _{1}(r)=\sqrt{\frac{32\pi N}{\left( 2\pi ^{2}-15\right) D^{3}}}\cos
^{2}\left( \frac{\pi r}{D}\right) H_{\mathrm{s}}\left( \frac{D}{2}-r\right) ,
\label{compact}
\end{equation}%
\begin{equation}
\phi _{2}(r)=\frac{\sqrt{N}}{\pi ^{4/3}\rho ^{3/2}}\exp \left( -\frac{r^{2}}{%
2\rho ^{2}}\right) ,  \label{Gaussian}
\end{equation}%
where $N$ is the total norm (\ref{total-norm}). Substituting these \textit{%
ans\"{a}tze} into Lagrangian (\ref{Lagrange}), the calculation can be
performed analytically (in particular, it is convenient to calculate the
double integral in Eq. (\ref{Lagrange}) using the coordinate transformation $%
\mathbf{r}-\mathbf{r}^{\prime }\equiv \mathbf{r}_{-},\mathbf{r}+\mathbf{r}%
^{\prime }\equiv \mathbf{r}_{+}$, taking into account the respective
Jacobian $1/8$). The results for the compacton (C) and Gaussian (G) are
\begin{equation}
L_{\mathrm{VA\text{-}C}}=-\frac{\mu }{2}N+\frac{23N}{D^{2}}+\frac{2gN^{2}}{%
D^{3}}-\frac{2N^{2}\kappa }{D},
\end{equation}%
\begin{equation}
L_{\mathrm{VA\text{-}G}}=-\frac{\mu }{2}N+\frac{3N}{4\rho ^{2}}+\frac{\sqrt{2%
}gN^{2}}{16\pi ^{3/2}\rho ^{3}}-\frac{\kappa N^{2}}{\sqrt{2\pi }\rho }.
\end{equation}%
Then, values of variational parameters $D$ and $\rho $ in \textit{ans\"{a}tze%
} (\ref{compact}) and (\ref{Gaussian}) are predicted by the Euler-Lagrange
equations $\partial L_{\mathrm{VA\text{-}C}}/\partial D=0,\partial L_{%
\mathrm{VA\text{-}G}}/\partial \rho =0$:
\begin{equation}
8\kappa ND^{2}-220D-31gN=0,  \label{D2}
\end{equation}%
\begin{equation}
\kappa N\rho ^{2}-4\rho -0.1gN=0.  \label{rho2}
\end{equation}

\begin{figure}[tbp]
{\includegraphics[width=2.5in]{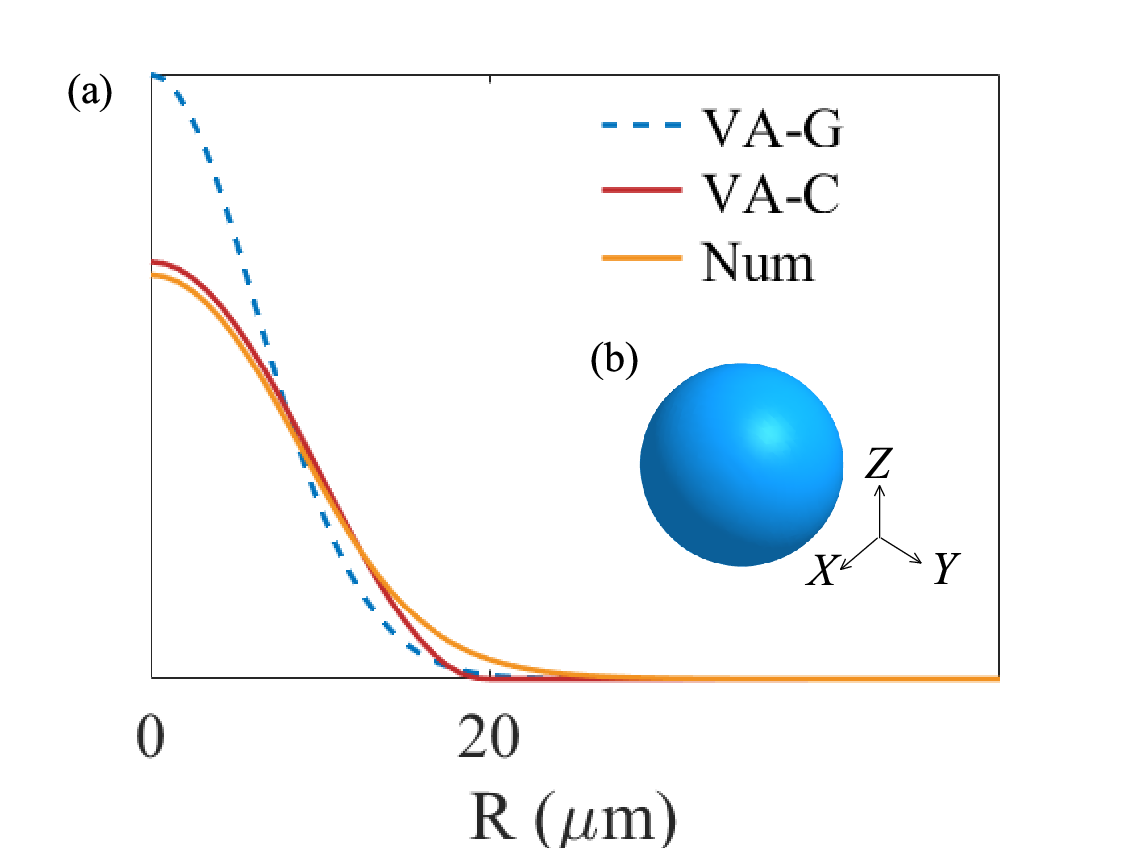}}
\caption{A typical example of the stable GS with parameters correspondong to
the condensate of $30000$ atoms, with physical parameters $a_{s}=100a_{0}$,
and $I=100\ \mathrm{MW}/\mathrm{cm}^{2}$ ($a_{0}$ is Bohr radius). The
corresponding dimensionless parameters are $g = 1.33$, and $\protect\kappa =
0.2$. (a) The blue dashed and red solid curves represents the VA prediction
based, respectively, on the Gaussian and compacton \textit{ans\"{a}tze} (%
\protect\ref{compact}) and (\protect\ref{Gaussian}), respectively. The
orange solid curve is the numerical solution for the same parameters. (b)
The density isosurface of the numerical solution. Direct simulation of the
perturbed evolution demonstrate that the GS is stable, at least, up to $%
T=100\ $ms (corresponding to the dimensionless time $t \approx 3000 $).}
\label{Example}
\end{figure}

In the case of the local self-repulsion, $g>0$ in Eq. (\ref{dime-less-GPE}),
each equation (\ref{D2}) and (\ref{rho2}) produces a single physically
relevant root:
\begin{equation}
D=\frac{220+\sqrt{979\kappa gN^{2}+48523}}{16\kappa N},
\end{equation}

\begin{equation}
\rho =\frac{3\sqrt{\pi }}{2\sqrt{2}\kappa N}+\sqrt{\frac{9\pi }{8\kappa
^{2}N^{2}}+\frac{3g}{8\pi \kappa }}.
\end{equation}%
Thus, once $N$, $\kappa $, and $g$ are fixed, the solution's width ($D$ for
the compacton and $\rho $ for the Gaussian) is predicted by the VA, which
subsequently defines their amplitudes, as per Eqs. (\ref{compact}) and (\ref%
{Gaussian}), respectively:%
\begin{equation}
\begin{split}
&A_{\mathrm{compacton}}=\sqrt{32\pi N/\left( 2\pi ^{2}-15\right) }D^{-3/2},
\\
&A_{\mathrm{Gaussian}}=\sqrt{N}\pi ^{-4/3}\rho ^{-3/2}.  \label{AA}
\end{split}%
\end{equation}

\begin{figure}[tbp]
    {\includegraphics[width=3.4in]{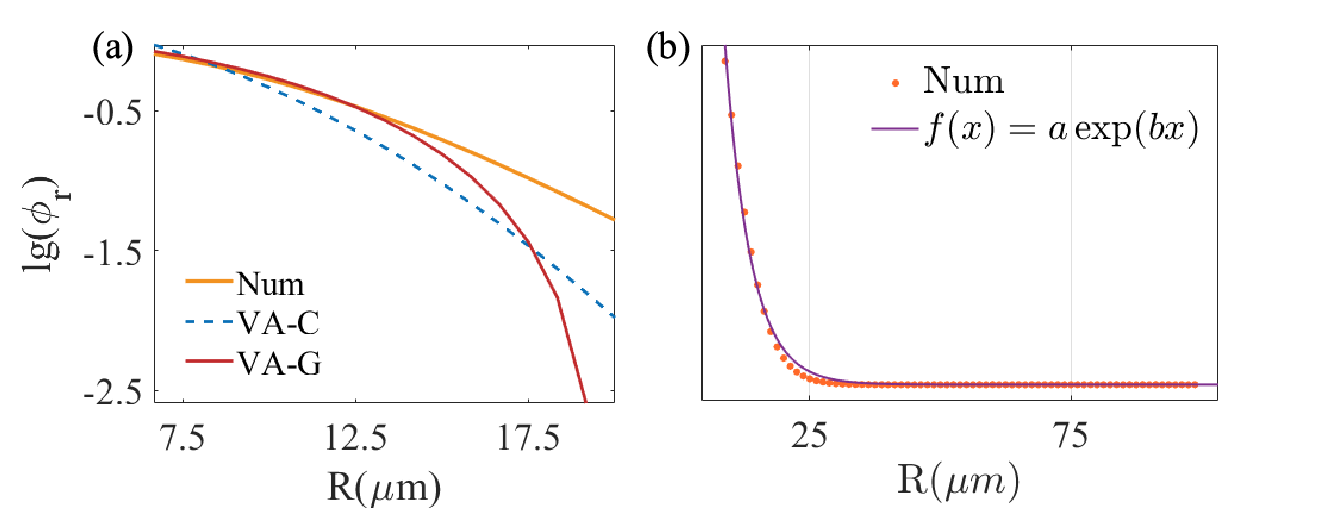}}
    \caption{Tail fitting: (a) displays the differences among the three wave 
    functions shown in Fig. \ref{Example} within the range \( R = 7.5 \sim 17.5 \, \mu\text{m} \). 
    The vertical axis is plotted on a logarithmic scale, denoted as \( \log_{10}(\phi)=\lg(\phi) \), 
    where \( \phi \) represents the corresponding wave function. Consistent with Fig. \ref{Example}, 
    the orange curve denotes the numerical solution, the blue curve corresponds to the variational 
    solution based on the Gaussian ansatz, and the red curve represents the variational solution 
    based on the compacton ansatz.  (b) shows a fitting of the numerical solution. The orange dots 
    represents the numerical result, while the purple curve is the fitted function. Here, we adopt 
    an exponential decay fitting of the form \( f(x) = a \exp(bx) \). In this case, the fitted 
    parameters are $a = 2.881$ and $b = -0.8895$.}
    \label{tail}
\end{figure}

\section{Stationary solutions}

\subsection{The ground state (GS)}

To explore characteristics of GS supported by the equation (\ref%
{dime-less-GPE}) of the SP type, we compare the VA-predicted solutions $\phi
_{1}(r)$ and $\phi _{2}(r)$ (the compacton and Gaussian approximations,
respectively) with a numerical solution $\phi (r)$ of Eq. (\ref{GP}) for the
same parameters, $\kappa $, $g$, and $N$.
The latter solution
was obtained as the GS of Eq. (\ref{dime-less-GPE}) using the
imaginary-time method \cite{yangNonlinear2010}, with the input
\begin{equation}
    \phi (r)= r^{\lvert S \rvert} \exp(-r^2/C + i S \theta),
    \label{GS-ansatz}
\end{equation}
where \(C>0\) is a constant, \( r = \sqrt{x^2 + y^2 + z^2} \),
\( \theta = \arctan(y/x) \), and $S = 0,1,2,...$ is the vorticity. The
results, shown in Fig. \ref{Example} (a), indicate that the compacton ansatz
provides a much better fit to the numerical solution. Figure \ref{Example}%
(b) displays its isotropic shape in the 3D space. Stability of the GSs was
tested by direct simulations of their perturbed evolution (not shown here in
detail), confirming that all GSs are stable modes, as it might be expected.

Nevertheless, the GS is not a genuine compacton. As shown in Fig. \ref{tail}(a), 
although the GS agrees well with the compacton solution in the central region, 
its tail does not exhibit the same rapid decay characteristic of compactons. 
In Fig. \ref{tail}(b), we fit the tail of the GS and find that it follows an 
exponential decay profile. Moreover, the decay rate is found to depend on 
the control parameter.

\begin{figure*}[tbp]
{\includegraphics[width=5in]{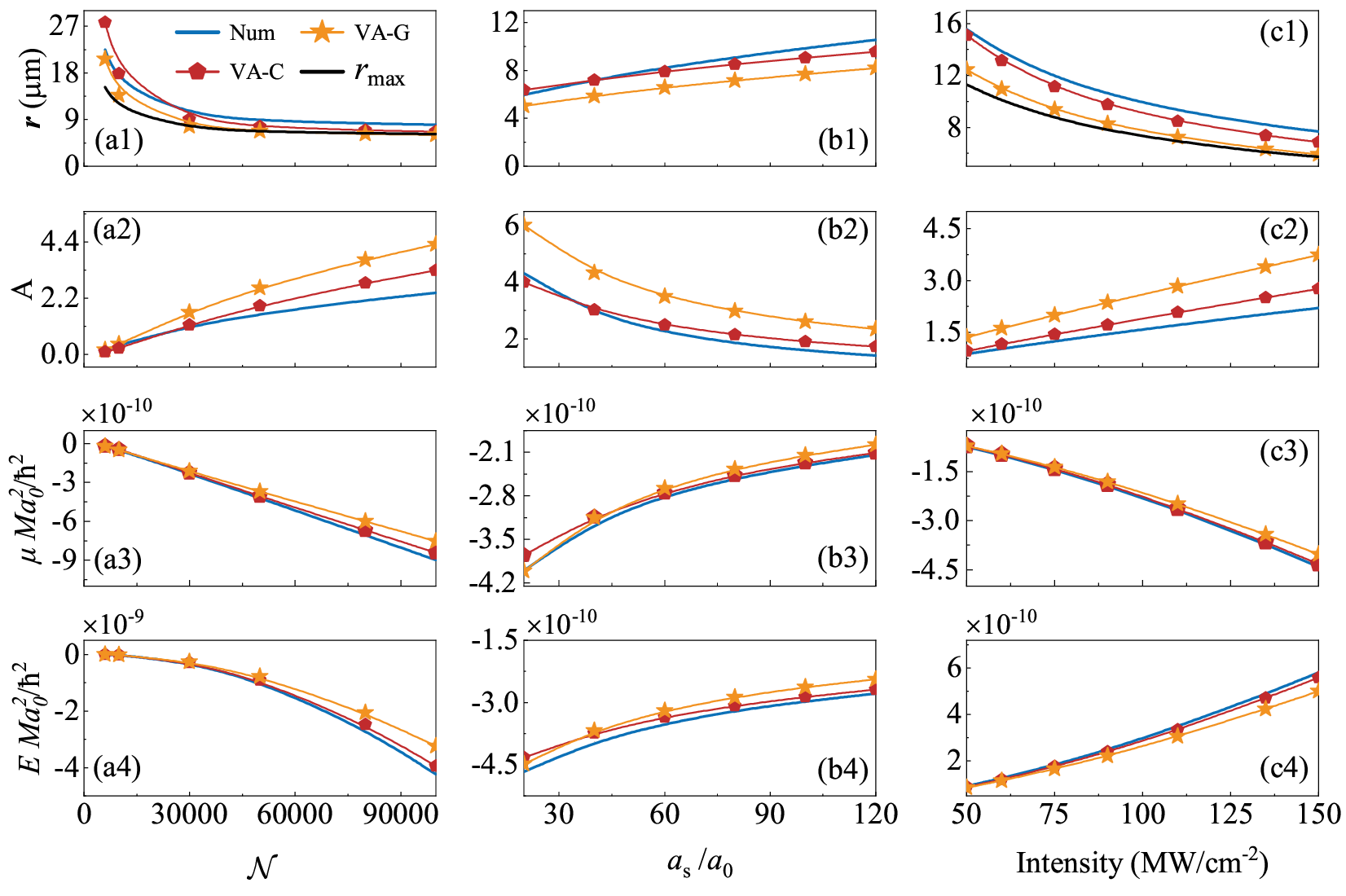}}
\caption{Radius $r$ (see Eq. (\protect\ref{r})), amplitude $A$ (see Eq. (%
\protect\ref{AA})), chemical potential $\protect\mu$ (see Eq. (\protect\ref%
{stationary-solution})) and energy $E$ (see Eq. (\protect\ref{energy})) vs.
the number of atoms $\mathcal{N}$ in the TSTMs (a1-a4),
scattering lngth $a_{s}$ (b1-b4), and intensity of the laser illumination
(c1-c4). The blue solid curves, chains of red pentagons, and orange
five-pointed stars represent the numerical results and VA predictions based
on the compacton and Gaussian ans\"{a}tze, respectively. The black solid
lines in (a1) and (c1) represent the value of $r_{\text{max}}\approx \protect%
\kappa N/(-2\protect\mu) $. In (a1-a4), we fix $a_{s}=100a_{0}$ and $I=100\
\mathrm{MW}/\mathrm{cm}^{2}$ (corresponding to $g = 1.33$ and $\protect%
\kappa = 0.2$). In (b1-b4), $\mathcal{N}=30000$ and $I=100\ \mathrm{MW}/%
\mathrm{cm}^{2}$ are kept constant. In (c1-c4), the parameters are fixed as $%
\mathcal{N}=30000$ and $a_{s}=100a_{0}$.}
\label{rAE}
\end{figure*}

To better illustrate the proximity of the self-trapped states \ to
compactons, we compare the radius
\begin{equation}
r=\sqrt{\langle r^{2}\rangle },~\langle r^{2}\rangle \equiv \int r^{2}d%
\mathbf{r}|\varphi (\mathbf{r})|^{2},  \label{r}
\end{equation}
amplitudes (\ref{AA}), chemical potential $\mu$ and energy (\ref{energy}) of
the Gaussian (VA), compacton (VA), and numerical solutions. In Fig. \ref{rAE}
(a1-a4) and (b1-b4), it is shown that the radius, amplitude and chemical
potential of the VA compacton are very close to their counterparts for the
numerical solution, while the Gaussian demonstrates an essential
discrepancy. This is consistent with the results shown in Fig. \ref{Example}%
. The results presented in Fig. \ref{rAE} indicate that, in terms of width,
amplitude, chemical potential and energy, the VA compactons are indeed very
close to the numerical solutions. Further, Fig. \ref{rAE} (a1-a4)
demonstrates that there exists a minimum (threshold) value of $\mathcal{N}$,
below which (at $\mathcal{N}<6000$, in this figure), self-trapped states
cannot be formed. Fig. \ref{rAE} (c1-c4) indicate that, as the nonlocal
interaction strength, determined by the illumination intensity, increases,
the discrepancy between the VA Gaussian solution and the numerical one
becomes more evident, whereas the VA compacton solution consistently remains
in close agreement with the numerical results, in line with the above
conclusion.

In addition, we compared $r_{\text{max}}$ with the radius of the numerical
solution in Fig. 2 (a1) and (c1). From these two figures, it can be observed
that when $N$ and $\kappa$ are relatively small, there is a significant
difference between $r_{\text{max}}$ and the actual radius of the %
TSTM. However, as $N$ and $\kappa$ gradually increase,
particularly with an increase in $N$, the difference between $r_{\text{max}}$
and the radius of the TSTM decreases. Thus, it can be
inferred that, once $N$ is sufficiently large, the radius of the
TSTM becomes close to the predicted value
$r_{\text{max}}$. This result is consistent with the above asymptotic
analysis of self-trapped states, indicating that when $N$ is sufficiently
large, the gravitational interactions will cause the self-trapped state to
exhibit a sharper tail, leading it to form as a TSTM.

\begin{figure}[tbp]
{\includegraphics[width=3.7in]{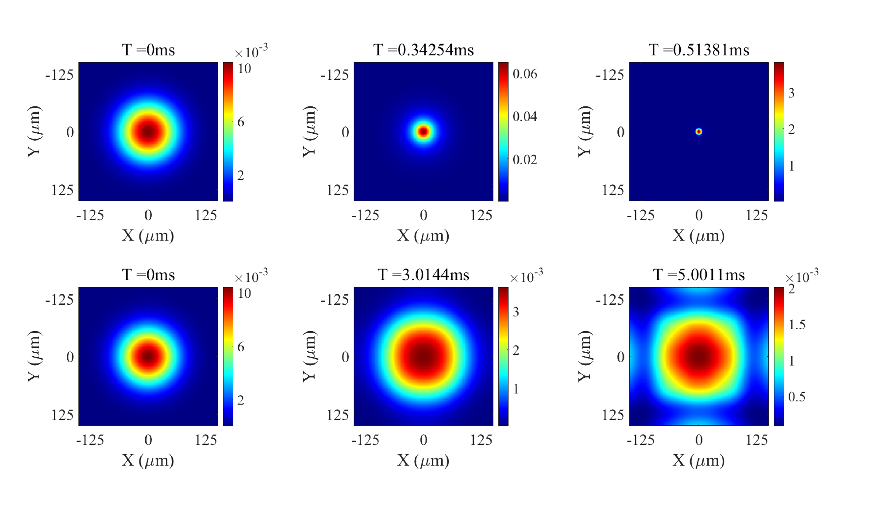}}
\caption{The top row displays the evolution of a loosely bound input into a
tightly bound state. The bottom row demonstrates that, in the absence of
long-range gravity-like interaction, the same input just spreads out (at the
late stage of the evolution, the isotropy of the expanding pattern is broken
by proximity to the domain's boundary). For this figure, the parameters are
chosen as: the total number of particles $\mathcal{N}=100000$, the
interaction strength $g=1.33$, and the long-range interaction strength $%
\protect\kappa=0.2$ and $\protect\kappa=0$, in the top and bottom rows,
respectively.}
\label{TSTMS-evolution}
\end{figure}

In our system, TSTMs emerge as a result of the intricate balance between
the gradient term (alias the kinetic energy term) and the interaction terms, which include
both local and nonlocal interactions. To further illustrate this point, we consider in Fig. \ref{TSTMS-evolution}
a loosely bound mode, given by \( \phi = A \exp(-R^2 / (2\sigma^2)) \) with
\( R = \sqrt{X^2 + Y^2 + Z^2} \) and a large width \( \sigma = 60 \, \mu m \), as an input
for the evolution governed by Eq. (\ref{dime-less-GPE}). Numerical simulations reveal that
this initial loosely bound mode rapidly self-traps into a tightly bound state, which remains
stable over an extended period (at least up to \( T = 70 \) ms in physical units), exhibiting
only minor intrinsic oscillations. These results provide clear evidence for the experimental
realization of tightly bound states and further confirm their robustness. Notably, if the \( \sim 1/R \)
nonlocal interaction term is removed from Eq. (\ref{dime-less-GPE}), the system fails to generate
any tightly bound state. This highlights the crucial role of the long-range gravity-like
interaction in supporting both the formation and stability of the tightly bound states.

It is worth noting that a compacton-type solution
was obtained in Ref. \cite{o2000bose}, using the Thomas-Fermi (TF) aooroximation, given by
\begin{equation}
    \Psi_{\text{TF-G}}(R) = \frac{\sqrt{N}}{2R_0} \sqrt{\frac{\sin(\pi R/R_0)}{R}}H_s(R_0 - R).
\end{equation}
Note that, in the framework of the
TF approximation, all solutions are compactons by definition.
The TF approximation reduces the original governing equations to
algebraic relations, neglecting the kinetic-energy terms and thus leading to solutions that
are nonzero only in regions where the potential and interaction terms are in balance, with
strict truncation elsewhere. Consequently, solutions produced by the TF
approximation always exhibit a compact support.

\begin{figure}[tbp]
{\includegraphics[width=3.4in]{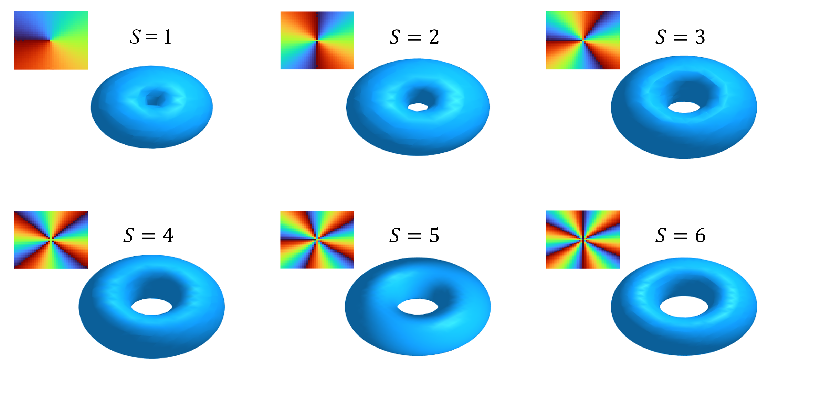}}
\caption{Density isosurfaces of stable vortices with topological charges $%
S=1\sim 3$ and $S=4\sim 6$ are plotted in the first and second rows,
respectively. The physical parameters are set as $a_{s}=100a_{0},I=100\
\mathrm{MW}/\mathrm{cm}^{2}$ and $\mathcal{N}=370000\ (S=1),400000\ (S=2),
435000\ (S=3),470000\ (S=4),500000\ (S=5),540000\ (S=6)$. In the upper left
corner of each density isosurface, the phase of the wave function is shown
in the $z=0$ cross-section. }
\label{vortex}
\end{figure}

\subsection{The vortex states}

As mentioned above, in a majority of other models admitting vortex solitons,
they are often unstable against spontaneous splitting, especially in the 3D
geometry \cite{malomed2022,kartashovFrontiers2019}. Therefore, it is
interesting to explore the TSTMs with embedded vorticity
and their stability in the present case. Because the
imaginary-time-integration method converges only to the GS solutions, we
here used the Newton conjugate gradient method (NCG) \cite{yangNCG2009} to
obtain vortex solutions.
For the vortex solution with
\( S = 1 \), the ansatz is given by Eq. (\ref{GS-ansatz}) with \( S = 1 \). In
contrast, for solutions with \( S > 1 \), we employ the modified ansatz
\begin{equation}
    \phi(r) = R'^{S} \exp(-R'^2/C + i S \theta),
\end{equation}
where \( R' = \sqrt{(r_{\bot} - R_0)^2 + z^2} \), \(r_{\bot}=\sqrt{x^2+y^2}\), and \( C > 0, R_0 > 0 \)
are constants. The modification of the ansatz for \( S > 1 \) is necessary because
such solutions exhibit a toroidal structure with a relatively large inner
radius $R_0$.

\begin{figure}[tbp]
{\includegraphics[width=3.4in]{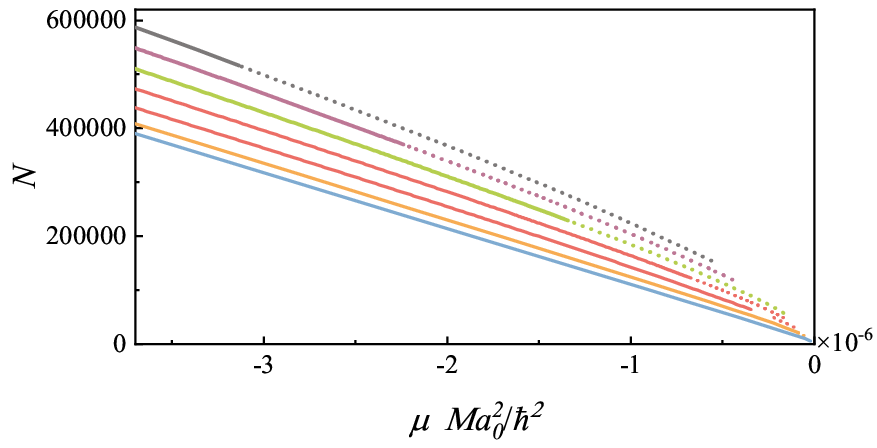}}
\caption{The total atom number $\mathcal{N}$ vs. chemical potential $\protect%
\mu $ is shown for the GS and vortex solutions. Solid and dotted lines
denote stable and unstable intervals, respectively. The curves from bottom
to top pertain to topological charges $S=0\sim 6$, in the increasing order.}
\label{vortex-mu-N}
\end{figure}

Figure \ref{vortex} displays density distributions in the vortex states with
the topological charge from $S=1$ to $6$, and the respective phase
distributions of the wave function in the $z=0$ plane, clearly revealing the
3D toroidal structure of these solutions.
The
inner hole in the vortex states expands as \(S\) increases. This feature is
common for vortices in continuous \cite{liu2022,li2018,huangdipolar2017,
huangExcited2018,zhongSelf-trapping2018,dongStableHigher2024,xuVortex2023}
and discrete \cite{zhaoDiscrete2022,chenHidden2024} systems. It is explained
by the fact that, as $S$ increases, the phase of the wavefunction varies
more rapidly in the azimuthal direction. Therefore, a larger inner radius of
the vortex is required to maintain the continuity of the wavefunction.

\begin{figure}[tbp]
{\includegraphics[width=3.4in]{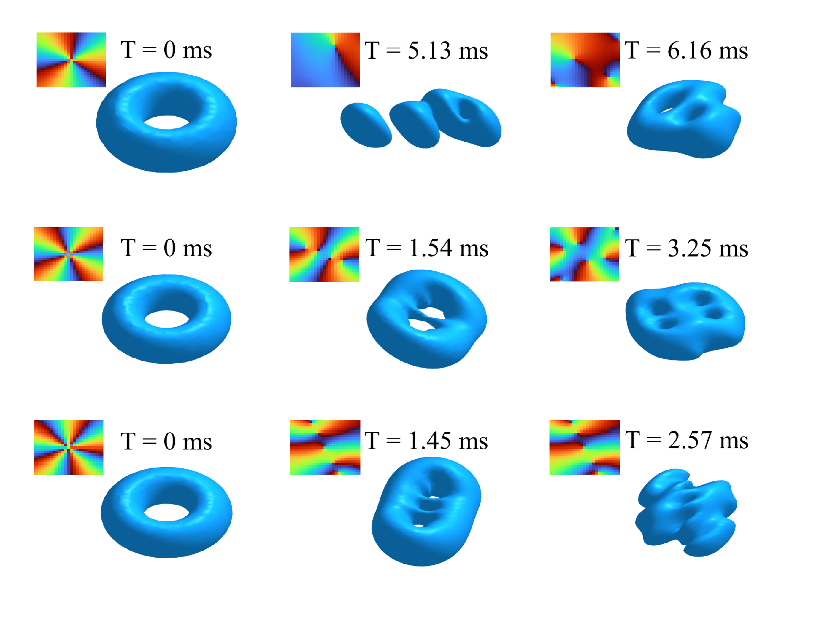}}
\caption{The unstable evolution of perturbed vortices for $S=3$ (first row),
$S=4$ (second row), and $S=5$ (third row) is shown, at various times, for
parameters $a_{s}=100a_{0},I=100\,\mathrm{MW}/\mathrm{cm}^{2}$ and $\mathcal{%
N}=60000$ for $S=3$, $\mathcal{N}=200000$ for $S=4$, $\mathcal{N}=220000$
for $S=5$.}
\label{unstab-vortex}
\end{figure}

Stability of the vortex states was verified, as in the case of GSs, by
simulations of their perturbed evolution in the framework of Eq. \ref%
{dime-less-GPE}. Figure \ref{vortex-mu-N} summarizes the results by means of
the $N(\mu )$ dependences for the GS and vortex states, with solid and
dotted segments indicating stable and unstable solutions, respectively.
As seen from Fig. \ref{vortex-mu-N}, when \( g \) is fixed, all vortex TSTMs
remain stable for large \( N \). The same result is valid for fixed
\( N \) and increasing \( g \). Thus, if \( g \) is fixed, a larger atomic number
\( N \) corresponds to stronger local repulsion, which favors the existence of stable vortex states.
These conclusions are generally similar to those reported in other models in the two-dimensional case
\cite{Kartashovstability,paredesVortex}. It is seen that, as the
topological charge increases, the stability interval gradually shrinks.
Thus, the $1/r$ long-range interaction is beneficial for the stable
existence of vortex states. As concerns the evolution of unstable vortex
states, it is displayed in Fig. \ref{unstab-vortex}, for $S=3,4,5$. It is
seen that unstable vortices fail to maintain their topological structure in
the course of the evolution, eventually decaying into GS.

\section{Dynamics}

\subsection{Orbiting bound states}

\begin{figure}[tbp]
{\includegraphics[width=3.4in]{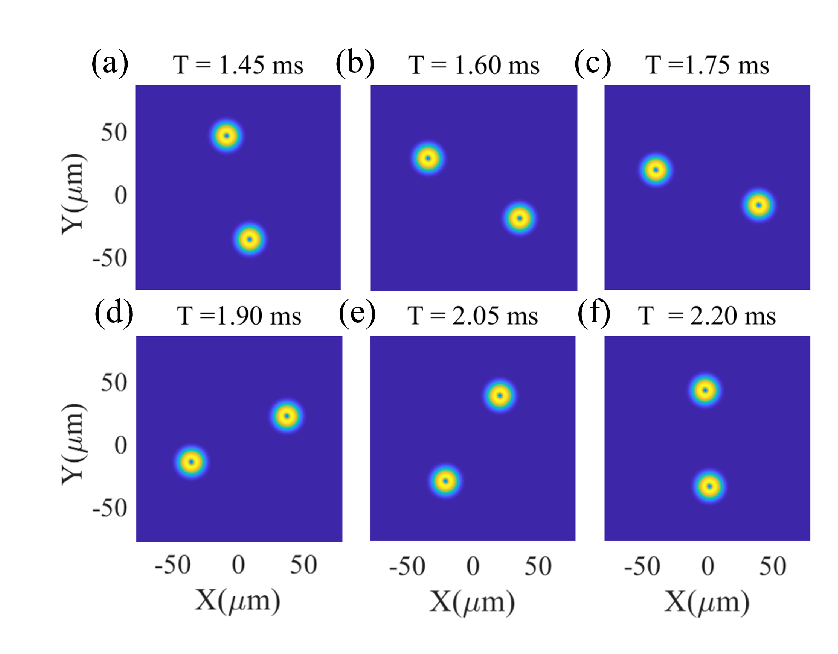}}
\caption{The stable rotation of the vortex-antivortex bound state ($%
S_{1,2}=\pm 1$), with physical parameters $a_{s}=100a_{0}$, $I=100\ \mathrm{%
MW/cm}^{2}$, and $\mathcal{N}=400000$. Panels (a)-(f) show the rotatiing
bound state at different times.}
\label{rotate}
\end{figure}

The next objective is to study two-TSTM dynamics in the model. In this
direction, we have found that two GSs, as well as two vortices, with
identical or opposite topological charges alike, can also stably rotate
around bound states. The dynamics is quite similar for the mutually orbiting
GSs or vortices. In Fig. \ref{rotate}, we address the rotating bound states
of two vortices with $S_{1,2}=\pm 1$ (the case of $S_{1}=S_{2}=1$ seems
nearly the same). The corresponding initial states were taken as%
\begin{equation}
\begin{split}
\phi (x,y,z,t=0)=& \varphi _{1}(x,y-y_{0},z,t=0)e^{-i\eta x} \\
& +\varphi _{2}(x,y+y_{0},z,t=0)e^{i\eta x},
\end{split}%
\end{equation}%
where $\varphi _{1}(x,y-y_{0},z)$ is the vortex with $S=1$, initially
centered at $y=-y_{0}$, and $\varphi _{2}(x,y+y_{0},z)$ is the vortex with $%
S=-1$ initially located at $y=y_{0}$. It is necessary to choose $y_{0}$
large enough, so that the two vortices are initially placed far from each
other, avoiding overlap. To set the pair in rotation, kicks of equal
magnitudes are applied to the upper and lower vortices in the negative and
positive $x$-direction, respectively. In Fig. \ref{rotate}(a)-(f), we
display the rotating bound states of the vortex-antivortex pair at different
times. The figure shows that, in the course of the presented time interval,
the two vortices rotate around each other by an angle close to 180$^{\circ }$%
. Although the distance between them somewhat varies during the rotation,
the bound state remains robust. While the figure displays only a half-period
of the orbiting motion, long simulations corroborate that the rotating bound
state remains stable, at least, until $T=50$ ms.

We have also tested rotating three- and four-body bound states, concluding
that they are always unstable, unlike the binary ones. In particular, a
bound state of three GSs can perform no more than two rotation periods,
which is followed by the merger or separation (not shown here in detail).

\begin{figure}[tbp]
{\includegraphics[width=3.4in]{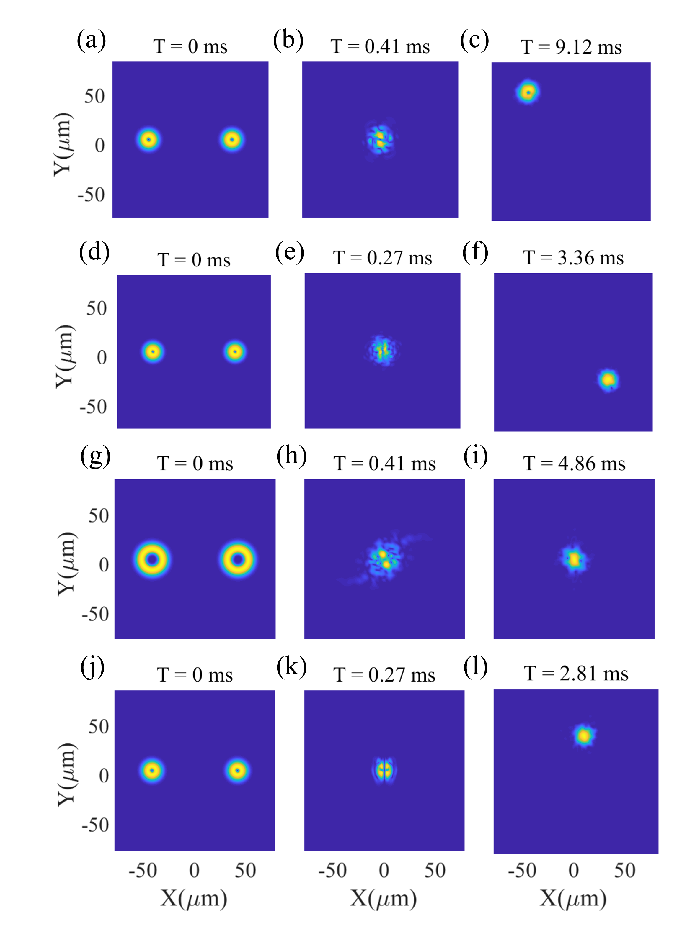}}
\caption{Collision dynamics:$a_s = 100 a_0 $, $I = 100\ \mathrm{MW}/\mathrm{%
cm}^2 $, $\protect\eta=0.1$, with $\mathcal{N} = 200000$ for $S = 1 $ (a-c)
and $S = -1 $ (j-l); $\mathcal{N} = 375000$ for $S = 1 $ (d-f); and $%
\mathcal{N} = 270000$ for $S = 4 $ (g-i). (a-c) illustrate the collision
between two $S = 1 $ vortices, where (b) depicts the moment of collision,
and (c) shows the formation of a single vortex moving in a specific
direction after the collision. Similarly, (d-f) also present the collision
of two $S = 1 $ vortices; however, due to the difference in atom numbers,
the outcome is a ground state that moves in a specific direction, as
indicated in (f). (g-i) demonstrate the collision between two $S = 4 $
vortices, characterized as a completely inelastic collision, with (i)
illustrating the transition to a ground state that remains stationary at the
collision point. Lastly, (j-l) depict the collision between a vortex and an
antivortex with $S_1 = -S_2 = 1 $; (k) represents the moment of collision,
while (l) shows the synthesis of a ground state that moves in a specific
direction following the collision. }
\label{collision-dynamics}
\end{figure}

\subsection{Collisions}

Apart from the formation of the rotational bound states,
it is also relevant to consider interaction
between TSTMs in the form of head-on collisions between them.
Previous studies primarily focused on the collisions between fundamental
solitons (with $S=0$), or between vortex states with $S=1$ \cite{Nikolaieva,Paredesinterference,
choicollision,bernalscalar}. In this section, we extend
the investigation by examining not only the collisions between
vortex TSTMs with \( S = 1 \), but also between ones with
(\( S > 1 \)), as well as vortex-antivortex collisions, with \( S = \pm 1 \).

Under the action of opposite kicks of size $\pm \eta $, two vortices
demonstrate head-on collisions between the vortices moving in the $x$
direction. The corresponding initial states were constructed as
\begin{equation}
\begin{split}
\phi (x,y,z,t=0)=& \varphi (x-x_{0},y,z,t=0)e^{-i\eta x} \\
& +\varphi (x+x_{0},y,z,t=0)e^{+i\eta x},
\end{split}%
\end{equation}%
where $\varphi (x\pm x_{0},y,z)$ represents the vortices with topological
charges $S_{1}$ and $S_{2}$ initially centered at $x=\pm x_{0}.$ Similar to
the case of rotation, it is necessary to choose a sufficiently large $x_{0}$
to prevent overlap between the vortices at $t=0$.

Fig. \ref{collision-dynamics} displays the collisions between two vortices
with $\eta=0.1$ and $S_{1}=S_{2}=1$, $S_{1}=S_{4}=4$ and $S_{1}=-S_{2}=1$ in
(a)-(f), (g)-(i), and (j)-(l), respectively. The difference between Fig. \ref%
{collision-dynamics}(a-c) and (d-f) is that in the former case they
represent the collision between the vortices with $S=1$ and $\mathcal{N}%
=200000$, while in the latter case the norm is essentially larger, $\mathcal{%
N}=375000$. In both cases, the collisions are completely inelastic, with the
difference that the \textquotedblleft lighter" vortices merge into a single
one with the the same topological charge, $S=1$, while the \textquotedblleft
heavier" vortices merge into a GS soliton with $S=0$. In either case, the
emerging fused soliton move in some direction (actually, a random one, which
is different in different realizations). Essentially the same outcomes are
produced by simulations of the collisions with the higher odd value of $%
S_{1}=S_{2}=3$. On the other hand, collisions between the vorticies carrying
even values of $S$, such as $S_{1,2}=4$, displayed in Figs. \ref%
{collision-dynamics}(g)-(i), leads to the merger into a single GS soliton
(with $S=0$), which remains quiescent thereafter.

In Fig. \ref{collision-dynamics}(j)-(l), the outcome of the
vortex-antivortex collision, with $S_{1,2}=\pm 1$, is quite
natural, \textit{viz}., the merger into a GS soliton. Somewhat similar to
the collision between the vortices with $S_{1,2}=1$, which is shown in Fig. %
\ref{collision-dynamics}(d)-(f), the emerging fused GS soliton is moving
aside.

Note that in Fig. \ref{collision-dynamics} the total angular momentum of the
interacting vortices is conserved in the collisions shown in panels (a-c)
and (j-l), while it is not conserved in the cases represented by panels
(d-f) and (g-i). The reason for the apparent non-conservation of the angular
momentum in the latter case is that, in the course of collision, a
considerable part of the angular momentum is radiated away with
small-amplitude waves, and eventually lost while hitting edges of the
integration domain.

The inelastic collisions are produced by the slow collisions, with
relatively small values of kick $\eta $. On the other hand, fast collisions,
initiated by large values of $\eta $, are quasi-elastic, i.e., the colliding
solitons pass through each other. Such results were demonstrated in
simulations performed for fast collisions between vortex solitons with $%
S_{1,2}=1$ in the framework of the 3D SP system in Ref. \cite{Nikolaieva}. A
threshold value of $\eta$, which a boundary between the elastic and
inelastic collisions, is significantly affected by the atom number $\mathcal{%
N}$ and topological charge $S$. In particular, for $S=1$ and $\mathcal{N}%
=80000$, the threshold corresponds to $\eta \approx 1$.

To conclude this section, it is relevant to mention that simulations of collisions between genuine
compactons performed in the framework of the respective real nonlinear equations, reveal creation of
small-amplitude compacton-anticompacton pairs \cite{rosenau1993compactons}. No similar effect is produced by the
present system, which stresses its difference from those which give rise to genuine compactons.

\section{Conclusion}

We have investigated self-trapped states in BEC with the gravity-like ($-1/r$%
) long-range interatomic interaction, that can be induced by the laser
illumination of the condensate. The system is described by the 3D GPE (%
Gross-Pitaevskii equation), which belong to the class of SP
(Schr\"{o}dinger-Poisson) models. Using the VA (variational approximation)
and systematically collected numerical results, we have found that the GPE
gives rise to the three-dimensional TSTMs (tightly self-trapped
modes). The VA, based on a compacton ansatz, produces results which are
very close to their numerical counterparts, while the Gaussian ansatz
produces less accurate predictions. The analytical solution for the decaying
tails of the self-trapped states explains the sharp edges
featured by TSTMs. It is also found that the self-trapping takes place
above a threshold value of the norm.
Real-time evolution demonstrates that under the influence of the \(-1/R\) nonlocal
interaction, TSTMs can spontaneously emerge from a loosely bound state. This finding
highlights that the long-range gravity-like interaction plays a key role in both
the formation and stabilization of the tightly bound state. In addition to
the GS (ground state), the system supports stable TSTMs
with embedded vorticity, which feature the 3D toroidal structure, with
topological charges up to $S=6$. Pairs of GSs, as well as vortex-vortex and
vortex-antivortex ones, give rise to stable rotating bound states. Head-on
collisions of the two vortices under small kick produce completely inelastic
outcomes, \textit{viz}., merger into a GS soliton that stays stationary at
the collision point or drifts aside, or the merger into a vortex state that
also moves aside.

As an extension of the current work, it may be interesting to add the
Lee-Huang-Yang (LHY) correction term to the underlying GPE, and explore
formation of self-trapped quantum droplets in the system, cf. Refs. \cite%
{petrovQuantum2015,petrovUltradilute2016}. A key feature of the quantum
droplets is their flat-top structure, resulting from the balance between the
cubic (mean-field) and LHY nonlinear terms \cite%
{luoAnew2020,timanpfanReviewQDs2020,bottcherNewstates2020}. In particular,
one may expect the existence of QCs with the flat-top intrinsic structure.

\begin{acknowledgments}
This work was supported by NNSFC (China) through Grants No. 12274077,
12475014, 11874112, 11905032, the Natural Science Foundation of Guangdong
province through Grant No. 2025A1515011128, 2024A1515030131, 2023A1515010770, the Research
Fund of Guangdong-Hong Kong-Macao Joint Laboratory for Intelligent
Micro-Nano Optoelectronic Technology through grant No.2020B1212030010. The
work of B.A.M. is supported, in part, by the Israel Science Foundation
through grant No. 1695/2022.
\end{acknowledgments}

\end{document}